\documentclass[12pt,preprint]{aastex}

\begin{document}

\title{A Multi-Wavelength Investigation of the Unidentified Gamma-Ray Source HESS\,J1708-410}

\author{Adam Van Etten\altaffilmark{1}, Stefan Funk\altaffilmark{1}, \& Jim Hinton\altaffilmark{2}}
\altaffiltext{1}{Department of Physics, Stanford University, Stanford, CA 94305}
\altaffiltext{2}{School of Physics \& Astronomy, University of Leeds, Leeds LS2 9JT, UK}
\email{ave@stanford.edu, sfunk@stanford.edu, j.a.hinton@leeds.ac.uk}

\begin{abstract}
  We report on recent XMM-Newton observations, archival radio continuum and CO
  data, and SED modeling of the unidentified Galactic plane source HESS\,J1708-410.  
  No significant extended X-ray emission
  is observed, and we place an upper limit of $3.2\times10^{-13} \rm
  \, erg \, cm^{-2} \, s^{-1}$ in the 2-4 keV range
  for the region of TeV emission.  Molonglo Galactic Plane Survey data
  is used to place an upper limit of 0.27 Jy at 843 MHz for the source, 
  with a 2.4 GHz limit of 0.4 Jy from the Parkes survey of the southern Galactic plane.
  $^{12}$CO (J 1$\rightarrow$0) data of this region
  indicates a plausible distance of 3 kpc for HESS\,J1708-410.
  SED modeling of both the H.E.S.S.\ detection
  and flux upper limits offer useful constraints on the
  emission mechanisms, magnetic field, injection spectrum, and ambient
  medium surrounding this source.

\end{abstract}


\section{Introduction}
Since beginning operation in 2004 the H.E.S.S. Cherenkov telescope array has proven a 
valuable tool in the investigation of extremely energetic phenomena, yielding
 $\sim 50$ very-high-energy (VHE) Galactic gamma-ray sources.
 Roughly half of these sources lack definitive counterparts at 
longer wavelengths, and a few lack \emph{any} plausible counterpart.
The firmly identified Galactic TeV sources consist of Supernova remnants (SNRs), 
Pulsar Wind Nebulae (PWNe) and high-mass X-ray binaries (HMXBs).
There are also plausible associations with young and massive stellar clusters \citep{aet07},
and dense molecular clouds \citep{aet06a}.
Those TeV sources with no low energy counterpart may 
be underluminous (at longer wavelengths) members of these source classes,
or may represent an entirely new class of TeV source.
Understanding these so called `dark accelerators' with no identified counterpart at other wavebands is one
of the more intriguing challenges of high energy astrophysics.

Gamma-rays in the VHE regime invariably arise from nonthermal particle populations,
though for a given source determining if this population is dominantly leptonic or hadronic 
often proves quite challenging.
For leptons in low density environments, inverse-Compton (IC)
upscattering of background photons off highly relativistic
electrons dominates, while bremsstrahlung becomes important only amongst
very dense surroundings.
For any scenario in which the gamma-ray emission arises from
relativistic electrons, significant X-ray synchrotron emission can be
expected, with a peak synchrotron flux comparable to the peak IC flux
for typical galactic magnetic field strengths and radiation densities. 
This synchrotron emission usually extends down to radio energies with 
emission from lower energy, $\cal{O}$(1 GeV), electrons. 
Yet if the electron spectrum within a source is truncated at low energies
a bright TeV source may be unaccompanied by a radio counterpart.  
Such a situation is plausible for PWNe, given that the lowest particle energy is determined primarily by the four velocity
of the pulsar wind upstream of the termination shock \citep{kc84b}
which is in turn linked to observable shock properties such as
synchrotron luminosity and shock radius.  A hard electron spectrum with index greater
than the canonical value of $-2$ might also result in an underluminous radio counterpart
 given the relative lack of low energy electrons.
Hadronic acceleration
results in the production of neutral pions via proton-proton
interactions, which decay to produce VHE gamma rays.  Proton
interactions also produce charged pions and hence secondary electrons
that in turn produce synchrotron radiation in the radio to X-ray range, 
albeit usually at a much lower level than in leptonic scenarios. Nevertheless, in
both scenarios, at least a weak lower energy counterpart is expected.
Probing this lower energy regime provides insights into the particle
population, ambient medium properties, and magnetic field strength.

HESS\,J1708$-$410 (hereafter J1708) is one of the relatively few VHE gamma-ray sources
for which \emph{no} plausible explanation has yet been put forward. 
It was discovered during the H.E.S.S.\ survey of the inner Galaxy, as described by \citep{aet06c}.  
Subsequent observations improved the
statistical significance to 11$\sigma$ over a total livetime of 39
hours \citep{aet08}.  J1708 is only slightly extended with respect to
the H.E.S.S.\ point spread function, consistent with a marginally elongated Gaussian of
approximately $0.06^{\circ} \times 0.08^{\circ}$.  \citet{aet08} report no
significant emission beyond 0.3$^{\circ}$, ruling out an association with
the nearby SNR G\,345.7$-$0.2.  
The nearest known pulsar (PSR J1707--4053) lies exterior to our XMM field of view
$\sim 15 \arcmin$ northwest of the H.E.S.S.\ centroid, and is
several orders of magnitude too weak ($\dot E \sim 4\times 10^{32} \, \rm erg \, s^{-1}$)
and too old ($\tau_c \sim 5 \times 10^{6}$ years) to conceivably power the TeV emission. 
\citet{aet08} fit the H.E.S.S.\ source spectrum over the
energy range of 0.50--60 TeV 
and an aperture of 0.3$^{\circ}$ to fully
enclose the source.  This yields a power-law index of 
$-2.5$ and a flux of $\approx$ 8$\times 10^{-12} \rm \,erg \, cm^{-2} \,
s^{-1}$.  An archival ASCA pointing of this region shows only a single
point source located over a degree from the H.E.S.S.\ source.  In addition, a
previous XMM-Newton exposure of  G\,345.7$-$0.2 revealed no X-ray emission at
the H.E.S.S.\ position, though this pointing placed J1708 close to the
edge of the field of view.  In this paper we report on new XMM-Newton
observations of this enigmatic VHE source.  

\section{Observations and Data Analysis}

J1708 was observed with the European Photon Imaging Cameras (EPIC)
aboard the XMM-Newton satellite in imaging mode with the medium filter
on 2006-08-25 through 2006-08-26.  
The calibration, data reduction,
and analysis rely upon the XMM-Newton Science Analysis Software
(SAS \footnote{The XMM-Newton SAS is developed and maintained by the Science
Operations Centre at the European Space Astronomy Centre and the Survey
Science Centre at the University of Leicester.}) version 8.0.0.
Following the standard procedure, the data were filtered to reduce
background caused by soft proton flares. 
As such flaring was minimal
in this observation, 24.0 ks of the total of 26.7 ks remain after filtering
for both the MOS1 and MOS2 camera.  For the PN camera the
27.0 ks were whittled down to 21.6 ks.  
The MOS cameras, with fewer chip gaps than the PN camera, are better
suited to imaging.  
Figure 1 shows a mosaic of the MOS1 and
MOS2 detectors revealing no sign of diffuse emission save for 
hard ($> 2$ keV) reflection contamination from an unknown bright source outside the field of view.  

\begin{figure}[h!]
\plottwo{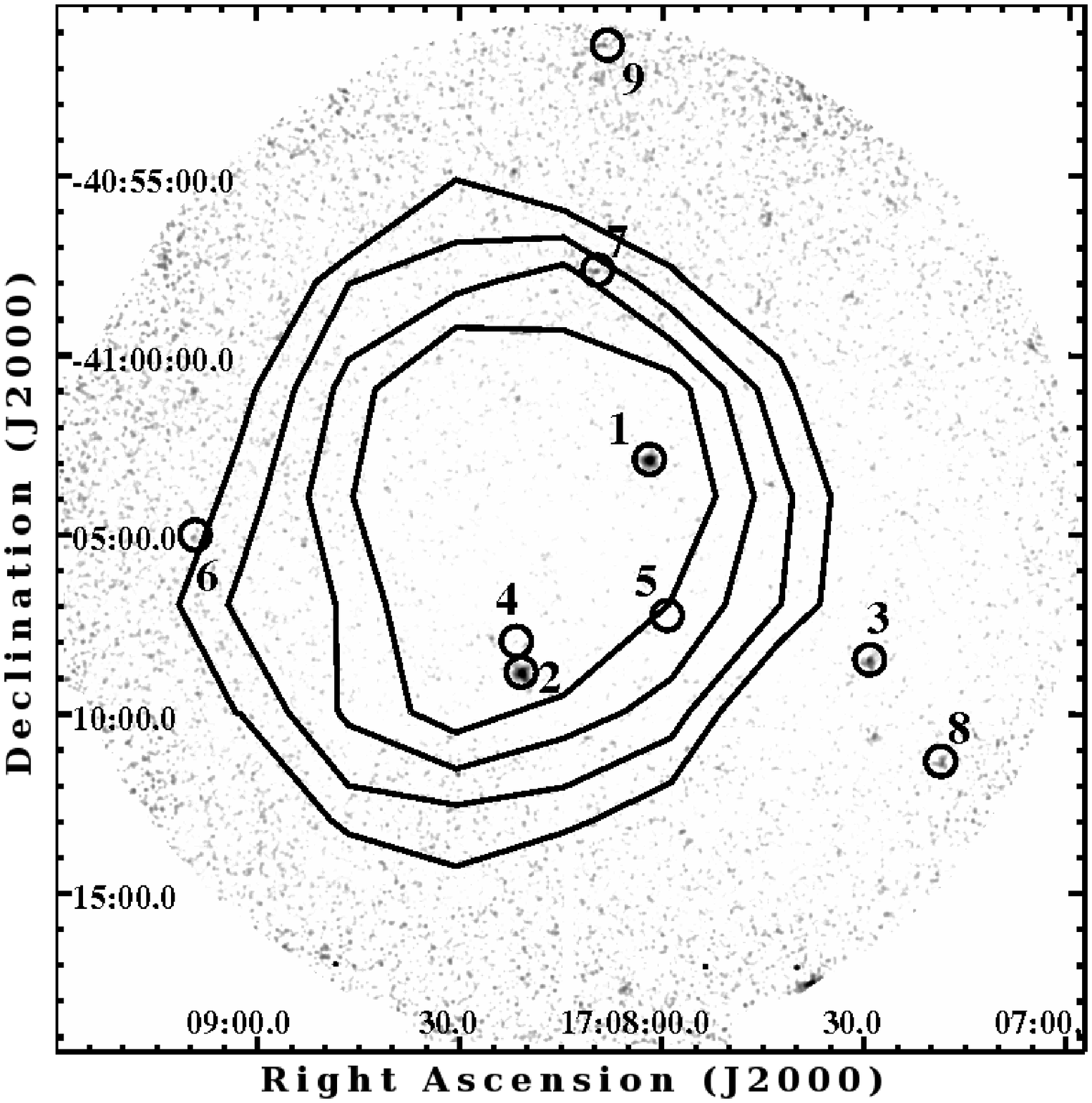}{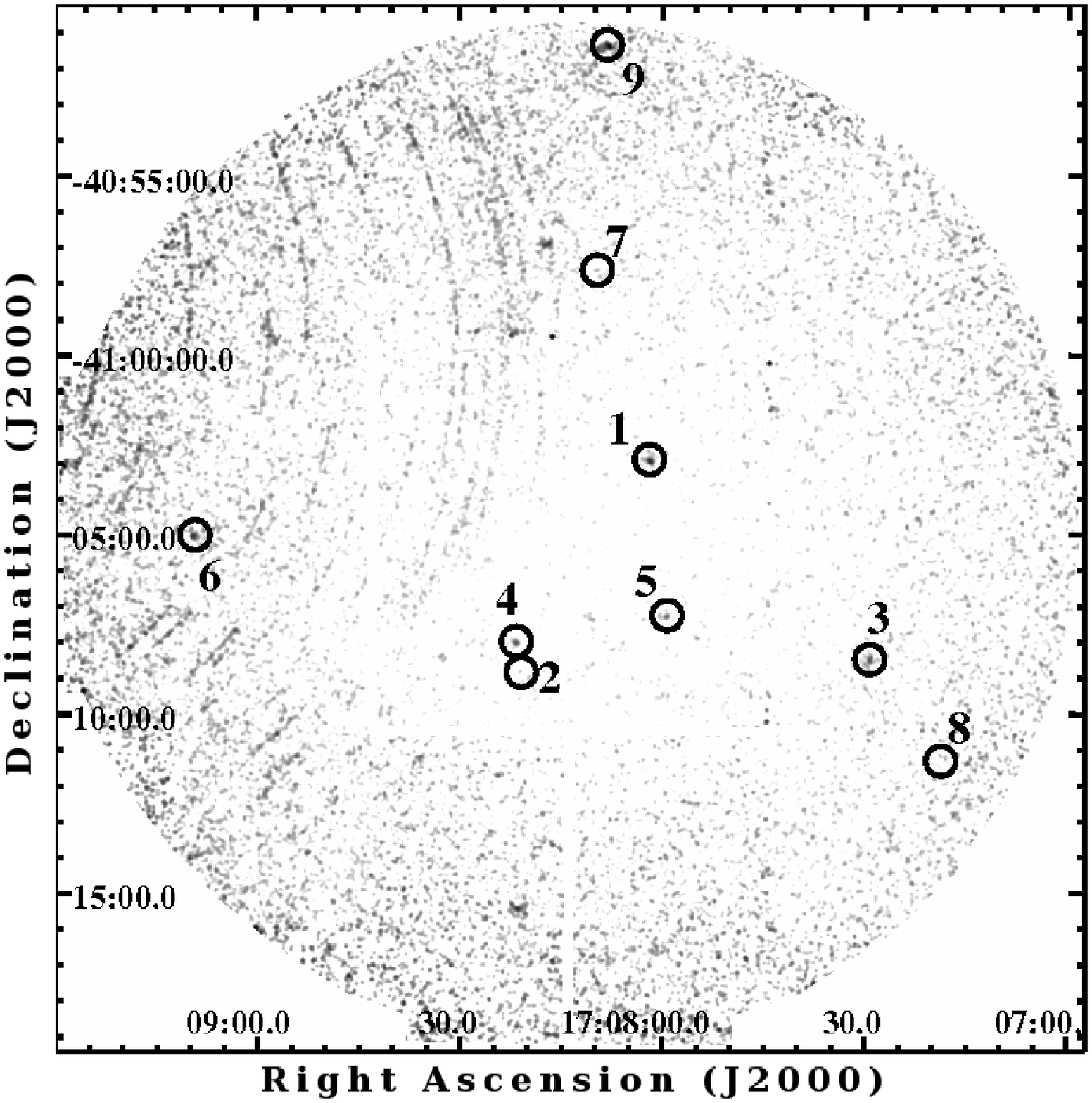}
\caption{
Left: EPIC MOS exposure corrected $0.3-2$ keV image with 
5.5$^{\prime\prime}$ Gaussian smoothing. 
The H.E.S.S.\  5, 6, 7, and 8 $\sigma$ contours are overlaid, and point sources indicated. 
Right: EPIC MOS exposure corrected $2-10$ keV image with 
5.5$^{\prime\prime}$ Gaussian smoothing exhibiting the reflection contamination.
Neither image indicates any sign of extended emission (save for contamination).
}
\end{figure}

\subsection{Counterparts Search}

Within the XMM-Newton field of view a number of point sources are detected.
No obvious extended emission is seen with the exception of contaminating
reflections from a nearby bright X-ray source at energies above 2~keV
(see Figure 1). Two ROSAT hard-band X-ray sources exist to the
northeast: 0.3$^{\circ}$ away lies J171011.5-405356, and slightly
farther away is LMXB 4U 1708-40, either of which could be responsible
for the reflection patterns.

We employ the SAS
function edetect$\_$chain for source detection and search in
two energy bands of $0.3-2$ keV and $2-10$ keV for sources 
with a probability $P < 10^{-13}$
of arising from random Poissonian fluctuations.  This rigorous cutoff threshold is necessary due to the 
aforementioned contaminating streaks.  
10 sources are detected, although we remove 1 source 
which lies directly along the arcs of reflection and has 0 counts in the soft band.  
The source
detection algorithm also attempts to determine source extension via a
Gaussian model, though all detections are consistent with a point source.
Table 1 lists the X-ray data.
We define the hardness ratio as: HR=$
(C_{hi}-C_{lo})/(C_{hi} + C_{lo}) $ where $C_{lo}$ and 
$C_{hi}$ are MOS counts in the $0.3-2$ keV and $2-10$ keV bands,
respectively.  The only source with a cataloged SIMBAD counterpart is source 2, \
which coincides with field star HD\,326944 of magnitude $\sim 10$ and type G0.  
Sources 1 and 3 lie within $1 \arcsec$
of USNO-B1.0 12th and 18th magnitude sources, respectively.   
Only sources 1 and 2 have sufficient counts to extract meaningful point source spectra, though fits to these sources are still accompanied by large errors. 
Source 1 is best matched with an absorbed power law, with 
hydrogen column density hardly constrained at
$n_H = 2.0_{-2.0}^{+2.7} \times 10^{21} \, \rm cm^{-2}$ (90\%
single parameter errors).  As such we fix $n_H$ at
$2.0 \times 10^{21} \, \rm cm^{-2}$ and measure an
 index of $2.9_{-0.4}^{+0.5}$ and unabsorbed 0.3--10 keV flux of
$1.3 \pm0.3 \times 10^{-13} \, \rm erg \, cm^{-2} \, s^{-1}$.
Source 2 (HD\,326944) is best modeled by an absorbed mekal plasma, with
$n_H$ again essentially unconstrained at
 $1.3_{-0.6}^{+5.4\times10^5} \times 10^{16} \, \rm cm^{-2}$.
With $n_H$ fixed at $1.0 \times 10^{16} \, \rm cm^{-2}$ we find a 
temperature of $kT = 0.49_{-0.10}^{+0.09} \, \rm keV$
and unabsorbed 0.3--10 keV flux
$4.2_{-0.7}^{+0.5} \times 10^{-14} \, \rm erg \, cm^{-2} \, s^{-1}$.

The 73 ms time resolution of
the PN camera allows searches for periodicities from a possible pulsar
candidate to be conducted; the MOS datasets have a time resolution of 2.6 s,
insufficient to search for a signal from a typical rotation powered
pulsar.  To this end, for each of the three brightest point sources we barycenter
the events, extract a lightcurve, and search for periodicities within
the PN dataset with the {\it Xronos} function {\it powspec}; no significant
periodicities are detected.

\begin{table}[!h]
\caption{Source Properties}
\tabletypesize=\scriptsize

\begin{tabular}{lccccc}
No. & R.A. & Dec. & Pos. Err.($\arcsec$)  & Counts & HR   \\
\hline

1 & $17:08:01.81$ & $-41:02:55.2$ & 0.37 &  $784.1\pm33.4$ &  $-0.65\pm0.035$   \\
2 & $17:08:20.81$ & $-41:08:51.2$ & 0.32 &  $624.4\pm29.5$ &  $-0.97\pm0.020$   \\
3 & $17:07:29.29$ & $-41:08:29.4$ & 0.46 &  $186.8\pm17.6$ &  $-0.26\pm0.096$   \\
4 & $17:08:21.68$ & $-41:07:59.5$ & 0.73 &  $72.7\pm12.1$  &  $+0.73\pm0.14  $   \\
5 & $17:07:59.38$ & $-41:07:16.0$ & 1.07 &  $57.0\pm11.1$  &  $+0.82\pm0.15  $   \\
6 & $17:09:09.23$ & $-41:05:00.8$ & 1.43 &  $59.1\pm11.1$  &  $+0.47\pm0.16$	\\
7 & $17:08:09.66$ & $-40:57:38.4$ & 1.15 &  $60.2\pm11.2$  &  $-0.79\pm0.17 $   \\
8 & $17:07:18.60$ & $-41:11:18.9$ & 1.46 &  $56.8\pm10.8$  &  $-0.85\pm0.18 $   \\
9 & $17:08:08.13$ & $-40:51:22.2$ & 0.93 &  $88.7\pm14.2$  &  $+0.57\pm0.13  $   \\

\hline
\end{tabular}
\label{srcprop}
\end{table}

\begin{figure}[h!]
\plotone{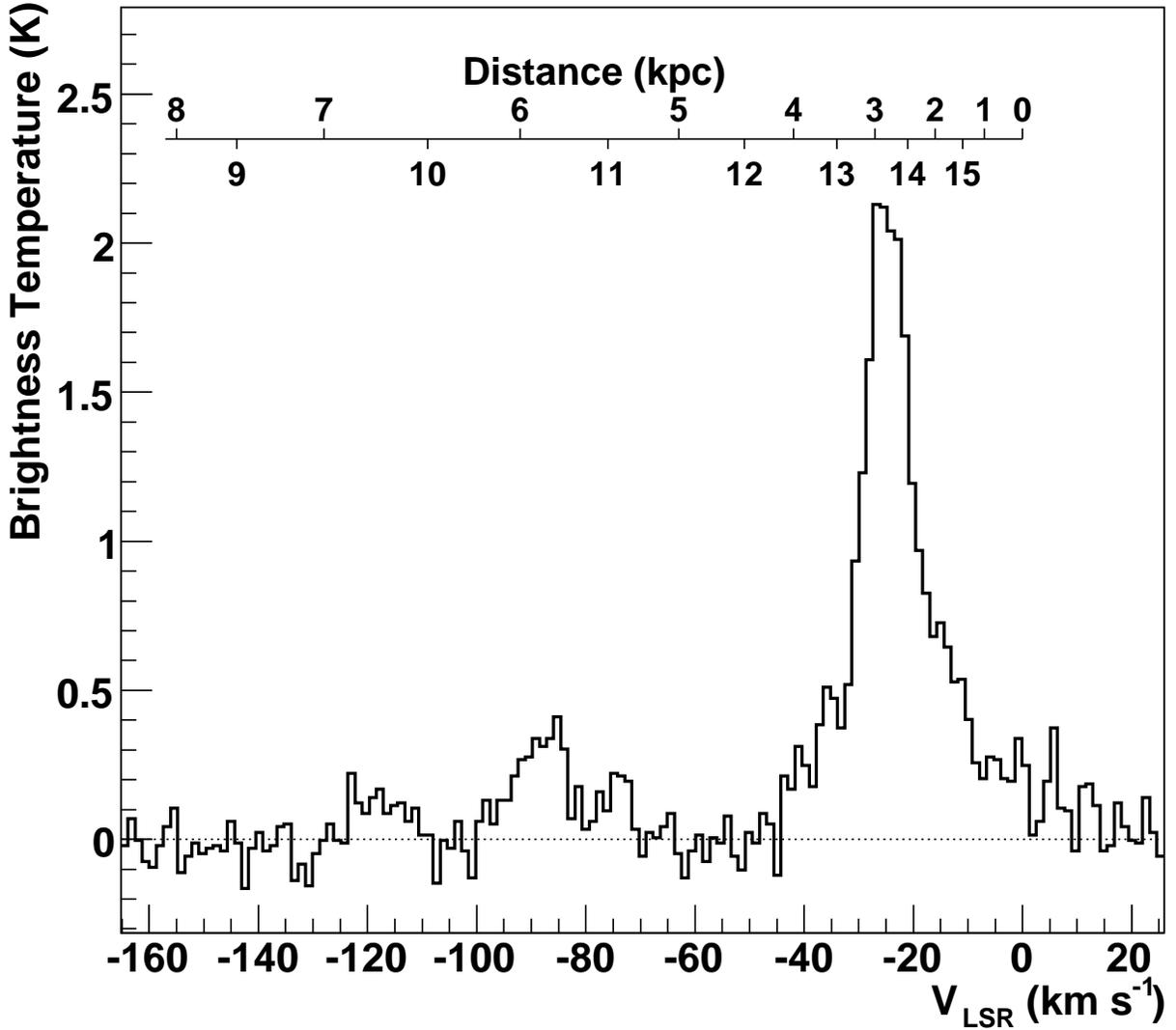}
\caption{
CO velocity profile in the vicinity of J1708 showing the single major molecular complex 
along the line of sight, with a probable distance of 3 kpc. The correspondence of
velocity to distance \citep[from the galactic rotation curve of][]{bb93} is inset.
}
\end{figure}

We analyzed $^{12}$CO (J 1$\rightarrow$0) data from the survey of
\citet{det01} to search for molecular cloud associations with J1708, 
constrain the target material density, and estimate 
the distance to J1708.  Figure 2
shows the CO velocity profile along the line of sight to J1708 in the
single $0.15^{\circ}\times 0.15^{\circ}$ pixel which covers the TeV source. A
single major molecular complex is visible in this profile with a
probable distance of $\approx 3 \pm 1$ kpc, adopting the Galactic rotation curve of
\citep{bb93} and assuming a near solution.  Effectively
all massive star formation occurs in molecular clouds, and all
known Galactic TeV sources are related to massive stars or their end
products (PWN, SNR, etc.). As such, this $3 \pm 1$ kpc distance (corresponding to the
near side of the Molecular Ring of the Milky Way) seems the most likely 
location for HESS\,J1708$-$410. We therefore assume a distance of 3 kpc in 
all subsequent discussions and scale
distances by $d_3 \equiv d/$(3 kpc).  A more speculative
alternative possibility is that J1708 is associated with the 
$\sim$0.1$^{\circ}$ cavity suggested by the Spitzer image of Figure 3, and 
comparable in size to the H.E.S.S.\ source.   
The region of low infrared
emission towards the center of J1708 also seems to correspond to a hole in
the nearby ($\sim$1 kpc) CO emission.  We could perhaps be looking at
a bubble blown by an SNR or stellar wind.  There is, however, no
evidence of a stellar cluster at the center of this bubble, which
undermines this hypothesis.

An upper limit on the density inside the source can be derived from
the column density towards the source and the observed source size
($\theta = 10'$ FWHM) $n \leq n_H / \theta \, D$. For the
measured molecular column around 3 kpc of $\approx 5 \times10^{21} \rm
\, cm^{-2}$ (24 K km with a conversion factor $X \approx 2 \times
10^{20} \, \rm cm^{-2}$ / (K km) ) 
this corresponds to $n \leq 200
\, d_3^{-1} \, \rm cm^{-3}$.  Note that the dominance of bremsstrahlung
over IC on the CMBR at photon energies of 1 TeV requires $n > 240 \,
\rm cm^{-3}$ for an $E^{-2}$ electron spectrum \citep{hah09}.  We
therefore neglect bremsstrahlung as an emission mechanism 
in the discussions below.

\begin{figure}[h!]
\plotone{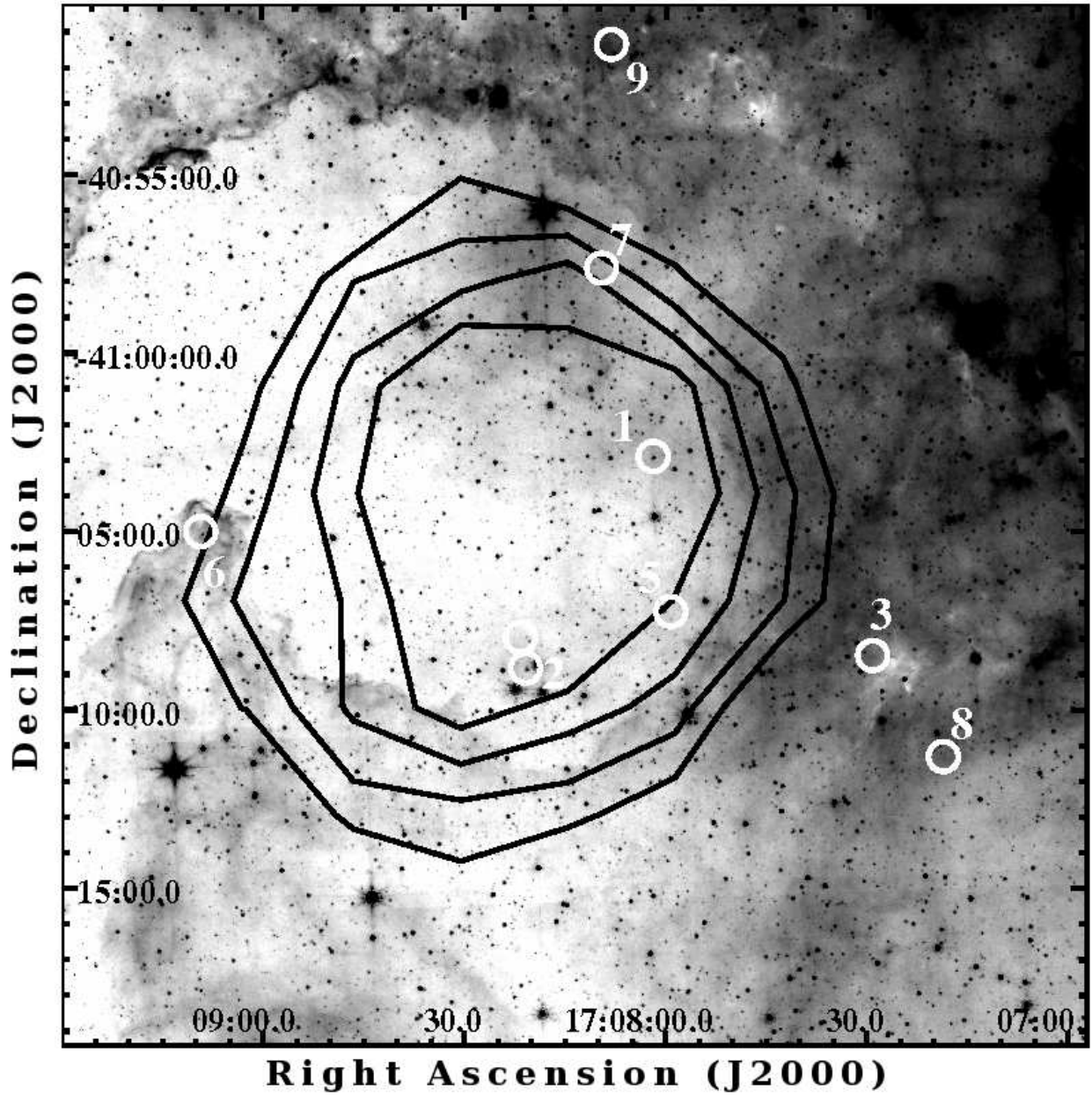}
\caption{
GLIMPSE Spitzer 8 $\mu$m image of the field of view of the XMM exposure, 
showing an apparent cavity coincident with J1708. H.E.S.S.\ significance 
contours and XMM source positions are overlaid.
}
\end{figure}

\subsection{Extended Source Spectral Analysis}

In the absence of any obvious extended X-ray emission from J1708, we fit the flux and use this
as an upper limit.  
To calculate this upper limit we define a region free of X-ray
point sources and scattered light contamination within the innermost (8$\sigma$) H.E.S.S.\  contour.  
This region encompasses 18\%
of the flux from the H.E.S.S.\ spectral integration area.  
We extract the spectrum from this region with the SAS
task evselect, and generate response files with the tasks arfgen and rmfgen.  The background
is taken from a region near the outer edge of the chip, exterior to the H.E.S.S.\ contours. Spectra
from both the MOS1 and MOS2 chips are fitted to a simple absorbed power law.  
This fit yields a hydrogen column density of
$2.1\pm{1.2}\times 10^{21} \rm \, cm^{-2}$, which can be compared to calculations
from HI maps of the total galactic column density.  
\citep{ket05} estimate a column of 1.34$\times10^{22} \rm \, cm^{-2}$,
and \citep{dl90} tabulate 1.65$\times10^{22} \rm \, cm^{-2}$.  
While one cannot make any strong statements with these estimates, it does at least indicate
that the fit hydrogen column density is consistent with a Galactic origin.

Scaling up the unabsorbed power law flux by the flux ratio of 1/0.18
(i.e. assuming matching X-ray and TeV morphology) provides a 90\% 
flux limit of $1.8\times10^{-12} \rm \, erg \, cm^{-2}
\, s^{-1}$ in the 0.3--10 keV range, and a limit of $3.2\times10^{-13}
\rm \, erg \, cm^{-2} \, s^{-1}$ between 2 and 4 keV.   
The 2--4 keV limit is
far less sensitive to the estimated column density to the source, 
so we quote this value in the following discussion.

While no deep radio images are available in this locale, we can 
extract upper limits from two major surveys of the southern Galactic plane.  
The first epoch Molonglo Galactic Plane Survey is a radio continuum survey 
at 843 MHz with a beam of 43$''$ $\times$ 65$''$ at the source position, and
root-mean-square sensitivity of $1-2 \rm \, mJy \, beam^{-1}$ 
\citep{get99}.
There is no sign of diffuse emission close to J1708 in the relevant Molonglo image;
one point-like source exists within the H.E.S.S.\  contours, but does not coincide with any of the XMM sources.
We calculate a conservative mean surface brightness of $3 \rm \, mJy \, beam^{-1}$ 
within the source region, and integrate over the 0.06$^{\circ} \times$ 0.08$^{\circ}$
 extent to derive a flux density upper limit of 270 mJy.
Higher frequency radio upper limits are available from the Parkes 2.4
GHz survey of the southern Galactic plane; images have a resolution of
10.4$'$ ($\approx$ the size of J1708) with rms noise of approximately
$12 \rm \, mJy \, beam^{-1}$ \citep{det95}.  We see no indication of a
source at the position of J1708, and derive an upper limit of $400 \rm
\, mJy \, beam^{-1}$.  As the source size is approximately the same as
the beam size, we place a flux density upper limit of 400 mJy at 2.4 GHz.

\section{Emission Scenarios}

The X-ray and radio flux limits place significant restrictions on
possible $\gamma$-ray emission mechanisms for J1708, which we explore by
modeling its spectral energy distribution (SED).  In Figures 4 and 5 we plot the 
SED of J1708 for both leptonic and hadronic
scenarios, showing the observed upper limits along with the
H.E.S.S.\ detection.  Also indicated is the one year sensitivity of the
Large Area Telescope aboard the Fermi Gamma-ray Space Telescope \citep{abdo09}.
To test these scenarios we apply a single-zone time-dependent
numerical model with constant injection luminosity. We inject a power law spectrum of relativistic
particles (either electrons or protons) with a high energy exponential cutoff, 
and then evolve this spectrum over time according to radiation losses from synchrotron and
IC (Klein-Nishina effects included) emission over a period of 1,000 - 
100,000 years. This age range is appropriate for most of the identified 
galactic plane TeV sources which are significantly extended.  
Many such sources are thought to be associated with the nebulae
surrounding Vela-like pulsars, which have characteristic spin-down ages of order 10 kyr.
More evolved sources are possible as well, although at least
in the leptonic scenario, cooling of VHE electrons limits lifetimes to 
$\sim$100 kyr as we shall discuss later. 
Injection (and evolution) occurs in time steps much smaller
than the assumed age.
More complex injection spectra, such as dual components, provide adequate
 fits to the data as well, though we see little justification for such
complications given the current limited data.
We consider a low density environment with density $n$ = 1 cm$^{-3}$
(although densities of up to $n$ = 200 cm$^{-3}$ are possible, 
as discussed in section 2.1),
and photon fields comprised of CMBR, far IR ($3 \times 10^{-3}$ eV,
with density 1 eV$\rm \, cm^{-3}$), and near IR (0.3 eV, with density
1 eV$\rm \, cm^{-3}$), appropriate for a typical inner Galaxy environment \citep{pms06}.

\subsection{Leptonic Gamma-Ray Emission Model}
 
The simplest reasonable injection model is an electron power law with a high energy cutoff.
We explore three source ages with this model: 1, 10, and 100 kyr. All ages yield comparable
fits to the VHE data, though source parameters must be varied. 
Enhanced cooling for older sources requires 
a greater value for the total energy and high energy cutoff,
as well as a lower magnetic field to allow
high energy electrons to survive and IC upscatter to TeV energies.  
For a 10 kyr source one can match the data with a total injection energy of
$5 \times 10^{47} \, d_3^2 $ erg, an electron power law of slope
$-1.8$, and an high energy exponential cutoff at
$14$ TeV.  The high energy cutoff is required to push the X-ray flux below the
new XMM limit; 14 TeV is the minimum allowed high energy cutoff
which still matches the H.E.S.S.\ data.  
One could fit the H.E.S.S.\ data
with a cutoff energy slightly greater than 14 TeV
(this would better match the highest energy H.E.S.S.\ points,
while worsening the match to the lower energy H.E.S.S.\ points), 
though this would require a lower
magnetic field due to the X-ray constraint on synchrotron radiation.  
For a 14 TeV cutoff the radio and X-ray limits impose an already 
rather stringent upper limit on the magnetic field of 4 $\mu$G for this 
spectral shape. 
A spectral index softer than
$-1.8$ is precluded by the radio upper limits.  Such a hard injection
spectrum would be remarkable, as a slope of $-1.8$ is not required
for any other galactic VHE source.  
For this electron spectrum, the synchrotron component
peaks in the UV, and IC emission peaks around 1 TeV.
Klein-Nishina effects suppress near IR scattering at the highest
energies, so only far IR and CMB photons contribute in the VHE regime, with far IR the dominant component. 
Figure 4 shows the SED in this scenario, as well as the SED for ages of 1 and 100 kyr.  
For a source age of 1 kyr and the same power law index
we require a cutoff at 8 TeV, an energy of $5 \times 10^{47} \, d_3^2 $ erg, and a maximum 
magnetic field of 4 $\mu$G. At this young age, the radio upper limits prove far more constraining
than the X-ray limits. A more aged source, at 100 kyr, has a total energy of $6 \times 10^{47} \, d_3^2 $ erg,
and requires a cutoff at 45 TeV and a very low magnetic field of 2 $\mu$G in order for energetic electrons to survive
the synchrotron and IC cooling process.  

We also
explore the consequences of implementing a low energy cutoff in the injection spectrum.  
As discussed in section 1, a low energy cutoff is expected for a PWN paradigm;
in the case of the Crab Nebula, 
\citep{kc84b} suggest a minimum injection energy of $\sim$ 1 TeV at the wind termination shock, 
while \citep{wet06} infer a minimum low-energy cutoff of 5 GeV for G\,359.95$-$0.04.
We inject an electron power law of slope $-2$, and a low energy cutoff of 10 GeV.
This low energy cutoff allows the softer electron spectrum to remain consistent
with the radio limits.
For an age of 10 kyr, we require
an exponential cutoff at 16 TeV, and a maximum magnetic field of 5 $\mu$G.
Matching the VHE data requires a
total of $6 \times 10^{47} \, d_3^2$ erg in electrons. Such a value is
reasonable for a PWN origin; one of the best studied galactic
broadband sources, the Vela X nebula surrounding the Vela pulsar, is
estimated to possess $\sim 10^{48}$ erg in the form of leptons
\citep{dJ07}.  
A 1 kyr aged source requires an exponential cutoff at 12 TeV, magnetic field
of 5 $\mu$G, and energy of $6 \times 10^{47}\, d_3^2$ erg.
Finally, a 100 kyr aged source requires an exponential cutoff at 60 TeV, magnetic field
of 2 $\mu$G, and energy of $8 \times 10^{47}\, d_3^2$ erg.
 
\begin{figure}[h!]
\plottwo{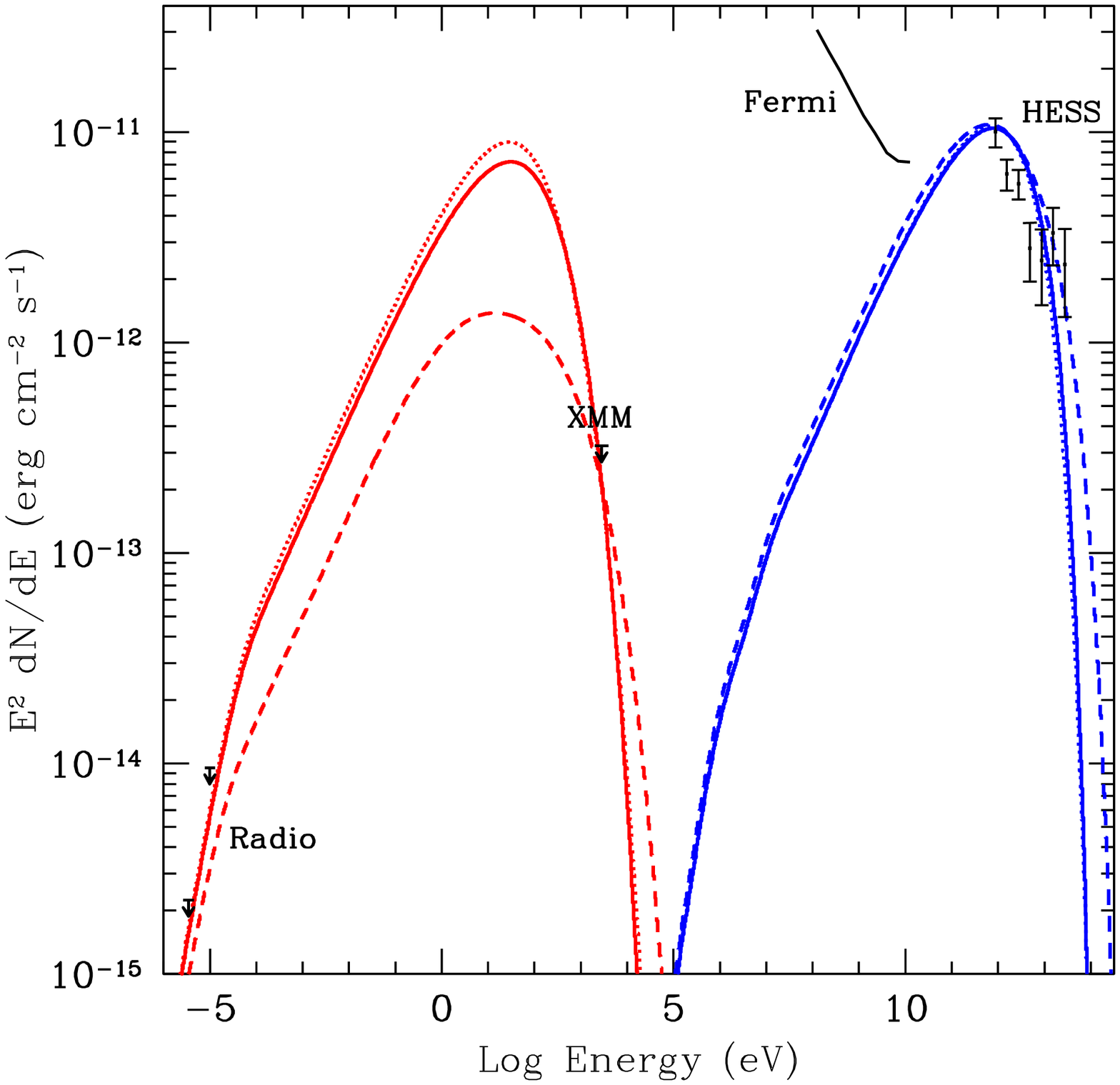}{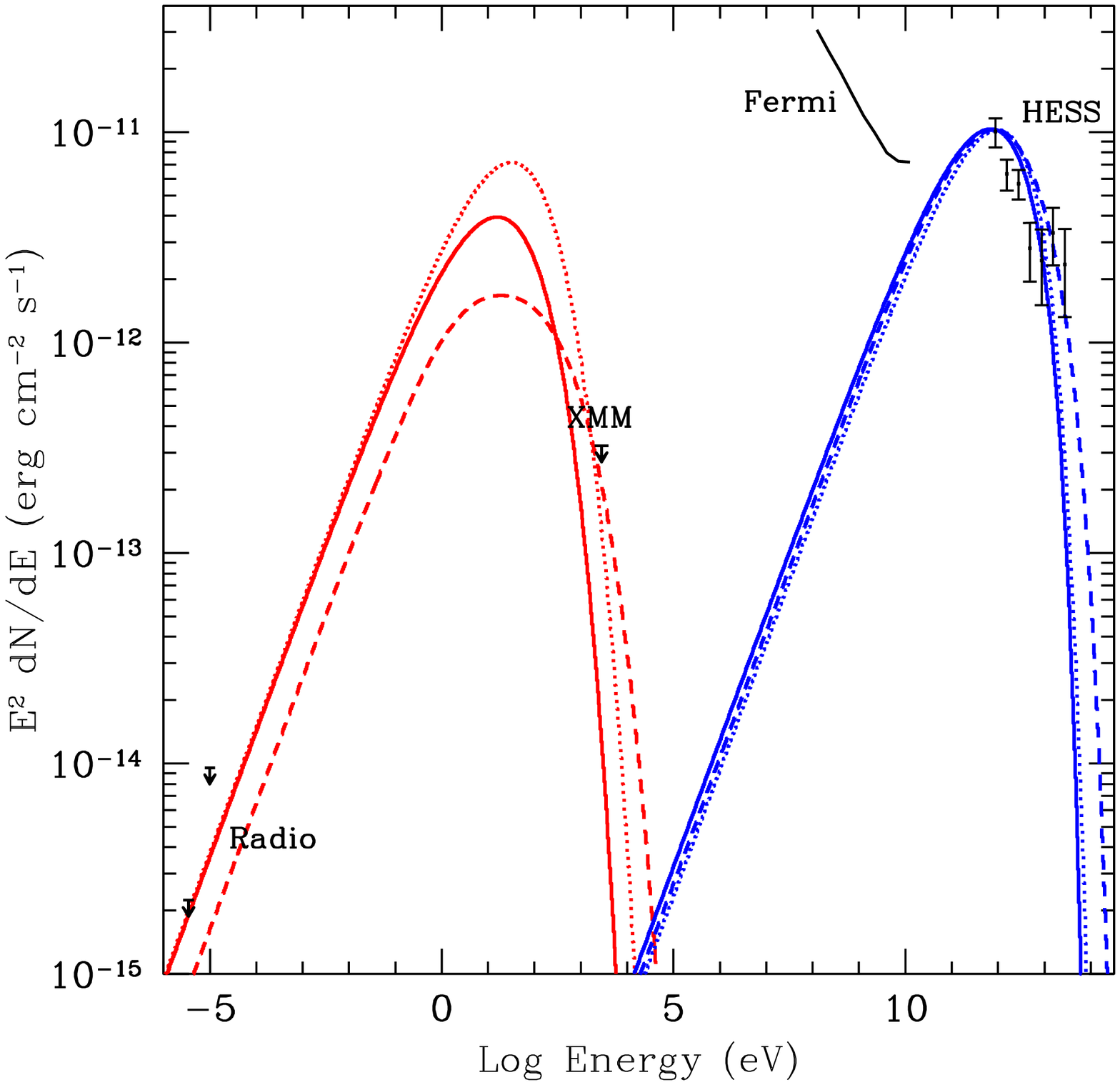}
\caption{
SED for leptonic scenario showing the synchrotron and total IC emission
for three source ages: 
Left: index of $-2$ and low energy cutoff of 10 GeV.  Solid denotes 1 kyr age (12 TeV cutoff, $B = 5 \, \mu \rm G$), 
dotted is 10 kyr (16 TeV cutoff, $B = 5 \, \mu \rm G$), and dashed is 100 kyr (60 TeV cutoff, $B = 2 \, \mu \rm G$).  
Right: index of $-1.8$ and no low energy cutoff.  Solid denotes 1 kyr age (8 TeV cutoff, $B = 4 \, \mu \rm G$), 
dotted is 10 kyr (14 TeV cutoff, $B = 4 \, \mu \rm G$), and dashed is 100 kyr (45 TeV cutoff, $B = 2 \, \mu \rm G$). 
}
\end{figure}

\subsection{Hadronic gamma-ray emission model}

A hadronic scenario relaxes the strictures imposed by the XMM upper limit, and instead one finds
the radio upper limits more constraining.
A proton power law with index $-2$ and 
cutoff at 50 TeV provides an adequate fit to the H.E.S.S.\  data, requiring energy
$E = 8\times10^{50} \, (n/1 \, \rm cm^{-3}) \,$$ d_3^2$ erg.
The timescale for pion production via p-p interactions of
$ \tau_{pp} \approx 1.5 \times 10^{8} \, (n/1 \, \rm cm^{-3})^{-1}$ years 
\citep{b70} is significantly greater than the expected age of the system, so the proton spectrum
is treated as static.   
For this proton spectrum, we calculate the photons from proton-proton interactions and 
subsequent $\pi^0$ and $\eta$-meson decay following \citep{ket06}.  Proton-proton interactions also yield $\pi^{\pm}$ mesons
which decay into secondary electrons, which we evolve in a 5 $\mu$G field.  The secondary electron
spectrum is a function of the age of the system, so we evolve over 1, 10, and 100 kyr, as shown in 
Figure 5. 
IC and synchrotron
fluxes from the resultant secondary electron spectrum are subsequently calculated, 
indicating that secondary IC 
is unimportant, while for high source ages one must 
take care to ensure the synchrotron component does not exceed the radio upper limits.  
We observe the $\pi^0$ gamma-rays as a broad peak in the GeV to TeV range, indicating that a Fermi 
detection would not be surprising.  

\begin{figure}[h!]
\plotone{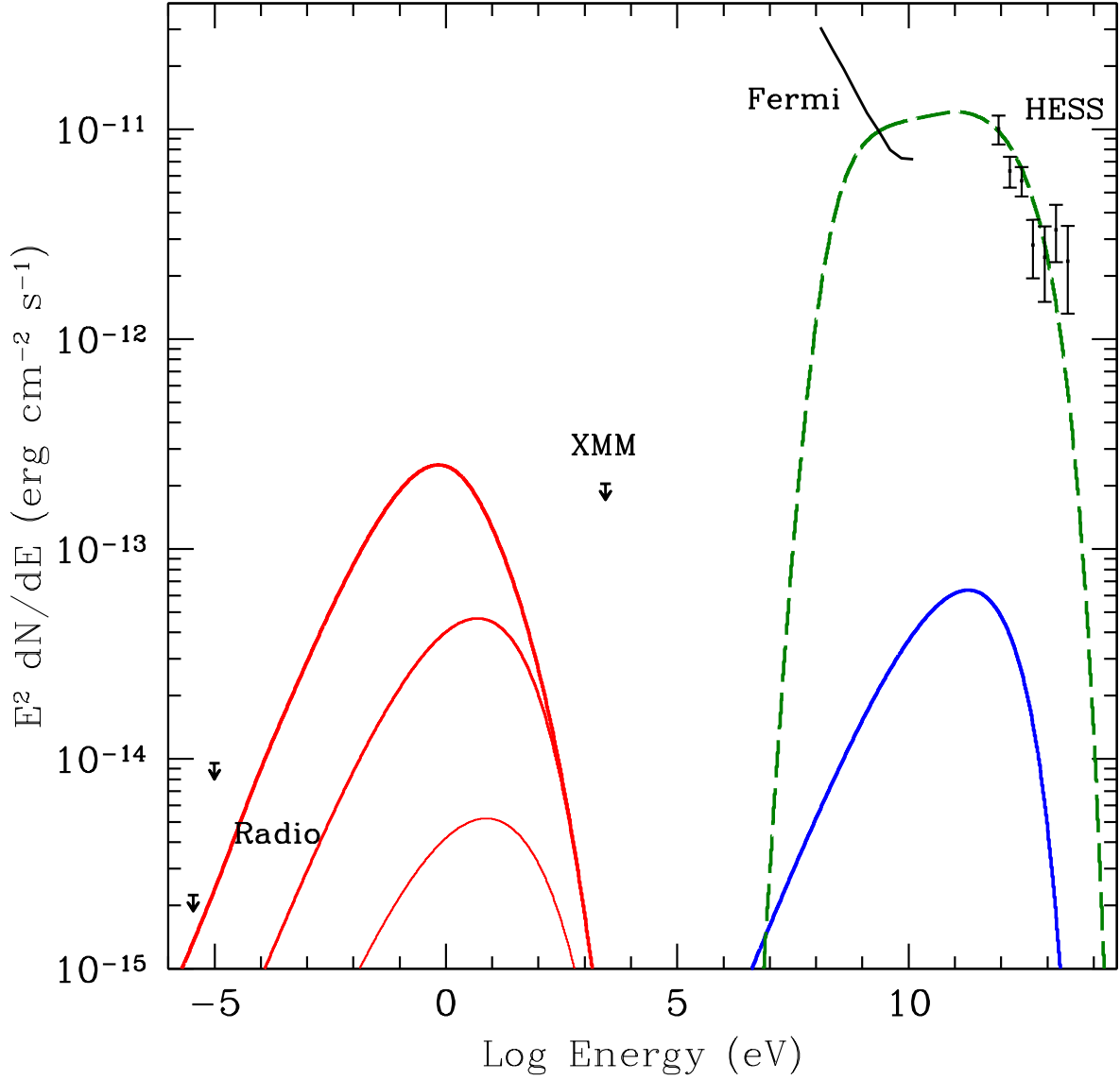}
\caption{
SED for a hadronic scenario:  $n=1$ cm$^{-3}$,  $B = 5 \mu$G, cutoff = 50 TeV.
Solid low energy lines indicate synchrotron emission for 1, 10, 100 kyr 
(bolder is older), higher energy solid line is IC emission for 10 kyr.  
The dashed line denotes gamma-rays from pion decay.  
A higher ambient particle density would decrease the energy
requirement in the form of hadrons, and correspondingly reduce the synchrotron and IC flux.  
}
\end{figure}

\section{Discussion}

\subsection{Leptonic Scenario}

The steep ($\Gamma = -2.5$) H.E.S.S.\ TeV spectral index could imply that
electrons responsible for this emission are cooled.  
As already discussed, in a leptonic scenario a high magnetic field is
precluded (given the low TeV to X-ray flux ratio), which
leaves a large age as the remaining cooling possibility.  
Indeed, for 
a source age of 1 kyr cooling is negligible, while at 10 kyr cooling is just
beginning to affect the electrons responsible for upscattering photons
to the VHE regime (which necessitates the slightly greater value for the high energy cutoff). 
The mean
electron energy required to IC scatter the CMB seed photons to TeV
energies ($E_{\rm TeV}$) is $E_e = \rm (18 TeV)$$E^{1/2}_{ \rm TeV}$, so the presence of
gamma-rays up to 50 TeV indicates electrons must be accelerated to
$\sim 100$ TeV levels.  For more modest 10 TeV electrons, the cooling
timescale for electrons IC scattering off the CMBR is $\tau_{IC} =
100 \, (E/10 \, \rm TeV)^{-1}$ kyr.  In regions of very low magnetic field
the IC timescale is comparable to the synchrotron cooling lifetime
$\tau_{sync} = 90 \, (E/10 \, \rm TeV)^{-1}$$(B/3 \, \mu \rm G)^{-2}$
kyr.  Regardless of whether electrons lose their energy primarily
through IC or synchrotron, based solely on cooling
timescales one might estimate J1708 to be as old as 100 kyr for very
low ambient magnetic fields.  
One can compare this age with estimates of the diffusion timescale.
A source subtending
$0.07^{\circ}$ has radius $R =  3.7 d_{3}$ pc. Then
for a cooling limited electron source with Bohm diffusion 
$t_{diff} = 19 \, d_{3}^2 \, (B/3 \, \mu \rm G)$$(E/10 \, \rm TeV)^{-1}$ kyr.  
This indicates a problem; it becomes difficult to confine the VHE electrons required
to power the H.E.S.S.\  source long enough for a cooling break to develop.  This argues
against a source age as high as 100 kyr, since
for the canonical electron power law index of $-2$ we found above that significant numbers
of electrons must persist up to 60 TeV to combat the high degree of cooling over such a lengthy period. 

Another possibility to explain the steep H.E.S.S.\ spectrum
is that there is a cutoff in the electron spectrum not far above
the H.E.S.S. range (at some tens of TeV) corresponding to either a maximum in the acceleration energy
or the escape of the highest energy electrons.  
A truncated accelerator provides a simple
explanation, but even if the accelerator persists up to the PeV range, 100 TeV electrons
tend to escape in 2 kyr or faster for Bohm diffusion.  
The SED fits indicate that a source aged 10 kyr will suffer some cooling 
due to the low magnetic field required by the X-ray limit, but the location
of the IC peak still depends critically on the value of the high energy cutoff.  
A cut-off, rather than fully cooled, distribution of electrons
therefore seems the most natural explanation for the H.E.S.S.\  data in the leptonic scenario, 
justifying our assumed injection spectrum of a power law with an exponential cutoff.  Such a cutoff
also argues against an age of 100 kyr, noted above.  

For an age of 10 kyr one class of potential counterparts, SNRs, could
easily occupy the volume assumed for a 3 kpc distance, since this
implies a very reasonable mean 
expansion velocity of 
$360 \, d_3 \, (\theta/0.07^{\circ}) \, (T/10 \, {\rm kyr})^{-1} \rm \, km \, s^{-1}$.  
Middle aged pulsars and their corresponding wind nebulae offer another
possibility.  The reverse shock from the parent SNR is expected to reach 
the PWN in $\sim7$ kyr \citep{randc84}, which (as in the case of the Vela PWN \citep{blondinetal01})
can displace the PWN far from the parent pulsar where the magnetic field can be
quite low and the the scenarios described above remain plausible.  Another
possibility stems from the high transverse velocity of pulsars, which
typically cross their parent SNR shells after $\sim$ 40 kyr
\citep{get06}.  At this late stage the pulsar has often escaped its
original wind bubble, leaving behind a relic PWN \citep{vdset04}.  
In this environment a lack of X-ray emission is not surprising, as the
ambient magnetic field can be very low, while VHE photons arise from
IC scattering of the relic electrons.

\subsection{Hadronic Scenario}

In the hadronic scenario the high energy cutoff at 50 TeV most likely stems from 
either the maximum accelerator energy or the 
diffusive escape of the highest energy particles. 
The Bohm diffusion timescale predicts
a loss of 50 TeV particles in $\approx 6$ kyr in a 5 $\mu$G field.  
The assumed source age of 10 kyr would therefore 
be reasonable. 
The new XMM limit places few
constraints on the model.  In the radio regime, however, the flux scales 
as the number of electrons $n_{e}\times B^{3/2}$, or since we assume constant 
injection of particles, flux scales as the age of the system $T \times B^{3/2}$;
for J1708 we find $(B/10 \, \mu {\rm G})^{3/2} \,(T/50 \rm \, kyr) < 1$.  
Though this limit is not very constraining, it does indicate that our
assumptions for age and magnetic field are reasonable.
The energy requirement of 
$E = 8\times10^{50} \, (n/1 \, \rm cm^{-3}) \,$$d_3^2$ erg is consistent with
a supernova origin for ambient densities of $\sim 10 \rm \, cm^{-3}$.

\section{Conclusions}

Independent of any model, the VHE luminosity of $L \approx  9 \times 10^{33} \, d_{3}^{2} \rm \, erg \, s^{-1}$
indicates that J1708 is a prodigious gamma-ray emitter.  
For comparison, the Vela X nebula surrounding
the Vela pulsar is observed to have gamma-ray luminosity of only 
$L_{0.6-65 \, \rm TeV} \approx 10^{33} \, \rm erg \, s^{-1}$ \citep{aet06b}.
This would imply a high gamma-ray luminosity if J1708 originates from a PWN.
We compare to the Vela PWN rather than the standard candle of VHE astronomy,
the Crab, since the lack of X-rays and hence low magnetic field
implies an older PWN than 1 kyr. 
The leptonic scenario requires fine-tuning of low and high energy cutoffs and
magnetic field to match the observed H.E.S.S.\  flux and lower energy upper 
limits.  
While hadronic injection requires less manipulation of injection spectrum and source properties, 
the required energy in protons of
$E_p = 8\times10^{50} \, (n/1 \, \rm cm^{-3}) \,$$d_3^2$ erg is quite high for low density environs.
Yet if the H.E.S.S.\ detection stems from protons interacting with a molecular clouds of density
$\sim 100 \rm \, cm^{-3}$, this energy requirement is significantly relaxed.  Regardless of injection species
the diffusion and cooling timescales for VHE particles hint at an age of 10-50 kyr for magnetic fields in 
the range of $3-5 \rm \, \mu G$.

The currently available data do not allow us to conclusively pin down the nature of the 
emitting particles in J1708.
A Fermi LAT detection would provide a significant boost to understanding this source,
and implies a hadron accelerator as is apparent in Figures 4 and 5. 
Deeper radio images would also help us to elucidate the processes at work in J1708.  Further lowering
the radio upper limit would constrain the spectral index of the
source, and/or the low energy cutoff for leptons; for hadrons the
product of age and $B^{3/2}$ would be constrained.
J1708 is particularly interesting among unidentified TeV sources as 
it has no lower energy candidate counterparts whatsoever and is rather compact.  
These features, combined with its high gamma-ray luminosity, could point to a very
different type of VHE system than those we currently know.

We thank NASA for supporting this research with grant NNX06AH66G.

\acknowledgements

{\it Facilities:} \facility{XMM-Newton}

\end{document}